# Diffusion modeling for Dip-pen Nanolithography

Apoorv Kulkarni
Graduate student, Michigan Technological University

## Abstract

The diffusion model for the dip pen nanolithography is similar to spreading an ink drop on a paper. Nanolithography uses Atomic Probe Microscope like probes to deliver the ink drop on to the substrate. The model considers the ink to be molecular. Some models have been developed considering fluid ink, some with thiol inks, particulate inks and polymer inks. The models consider the fluid ink with the added factors such as flow rate, detachment rate, effect due to relative humidity, multilayer flow, the ink drop, with meniscus etc., which makes the model complicated. The aim of this paper is to create a generalized diffusion model

## Introduction

Dip-pen Nanolithography (DPN) is a process of fabricating nanoscale structures. The process uses a technique where a tip similar to atomic force microscope is dipped in ink and the ink is transported to the substrate which produces the nanoscale structures. The structures that are formed are usually monolayer which are self-assembled. When the tip of the fabricator is brought in contact with the substrate a meniscus is formed. The ink molecules are diffused from the tip to the substrate [1]. The general idea of dip pen nanolithography is descried in the figure

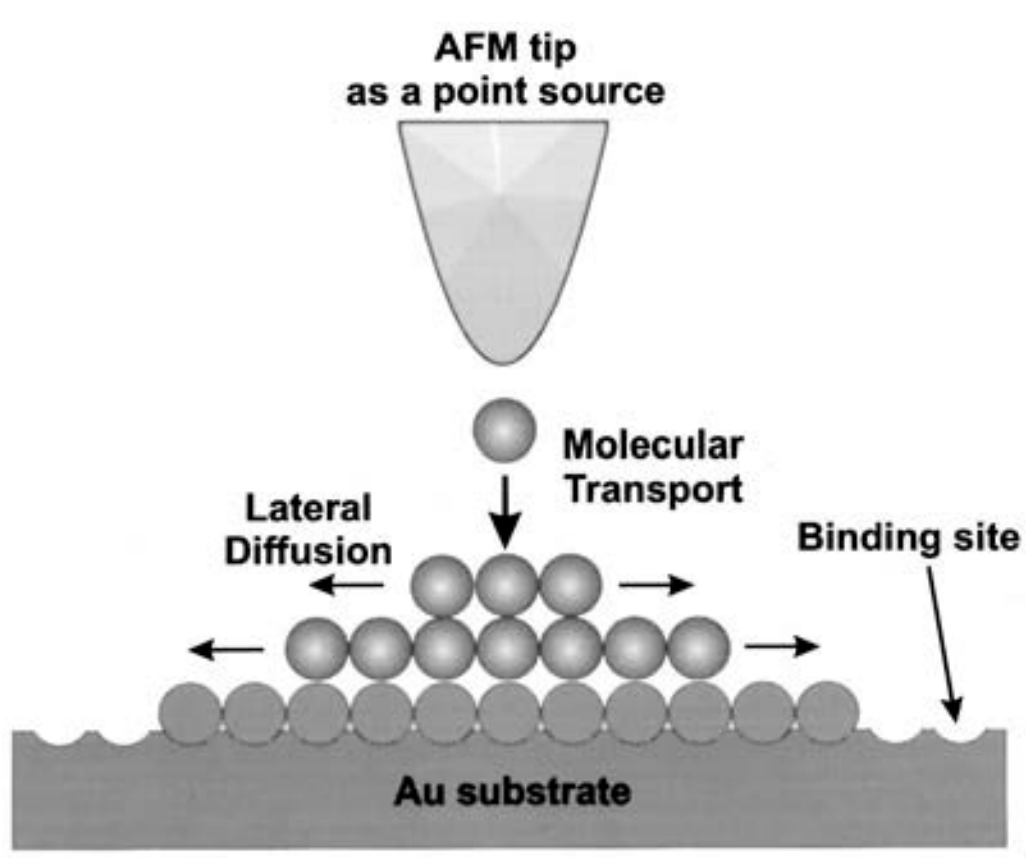

*Figure 1: Mechanism of dip pen nanolithography showing the transport of ink molecules from the tip to the substrate [2].*

The transport of the ink is controlled by many factors such as atmospheric conditions and surface of the water meniscus. The Dip-Pen Nanolithography (DPN) depends on the chemisorb of the substrate and the compact monolayer drop on the surface [1].The patterned features are governed by contact angle, contact time, retraction speed and volume of the ink at the tip. In this type of fabrication process, the inks do not form a chemical bond with the substrate. The phenomenon can be related to having just a liquid drop on the surface. [3]

## Existing models for Dip-Pen nanolithography

A model considered in the research paper “A surface diffusion model for Dip Pen Nanolithography line writing” considers a model for a continuous line with the inked tip moving and depositing the ink onto the substrate. They base their model on the 2D Fickian diffusion from the tip to the moving edge of the boundary [1].

$$\frac{\partial^2 C(x,y,t)}{\partial x^2} + \frac{\partial^2 C(x,y,t)}{\partial y^2} = \frac{1}{D} \times \frac{\partial C(x,y,t)}{\partial t}$$

Where they define a new co-ordinate system (ε,η) which is centered around the tip of the moving AFM, therefore the co-ordinates are obtained as follows $\varepsilon = x - Vt$ and $\eta = y$. Thus the equation is rewritten as follows.

$$\frac{\partial^2 C(\varepsilon,\eta,t)}{\partial \varepsilon^2} + \frac{\partial^2 C(\varepsilon,\eta,t)}{\partial \eta^2} = -\frac{V}{D} * \frac{\partial C(\varepsilon,\eta,t)}{\partial \varepsilon} + \frac{1}{D} \times \frac{\partial C(\varepsilon,\eta,t)}{\partial t}$$

Then axial symmetry is applied as a boundary condition considering the tip as the center of the cylindrical system.

$$\frac{1}{r}\frac{\partial}{\partial r}\left(r\frac{\partial C}{\partial r}\right) = 0C = -C_0 \frac{\ln(r / S)}{\ln(S / R)}$$

With $r^2 = \varepsilon^2 + \eta^2$ and boundary conditions,

$C = C_0$ @ r = R,

C = 0 @ r= S where S is the edge of the drop/line and R is the advancing radius.

The solution is given as

$$C = -C_0 \frac{\ln(r / S)}{\ln(S / R)}$$

The flux is then calculated as,

$$\bar{J} = \frac{2\pi D C_0}{\ln(S / R)}$$

By applying mass conservation, the relation between line width vs tip velocity and ink flow rate vs tip velocity are plot as follows [1].

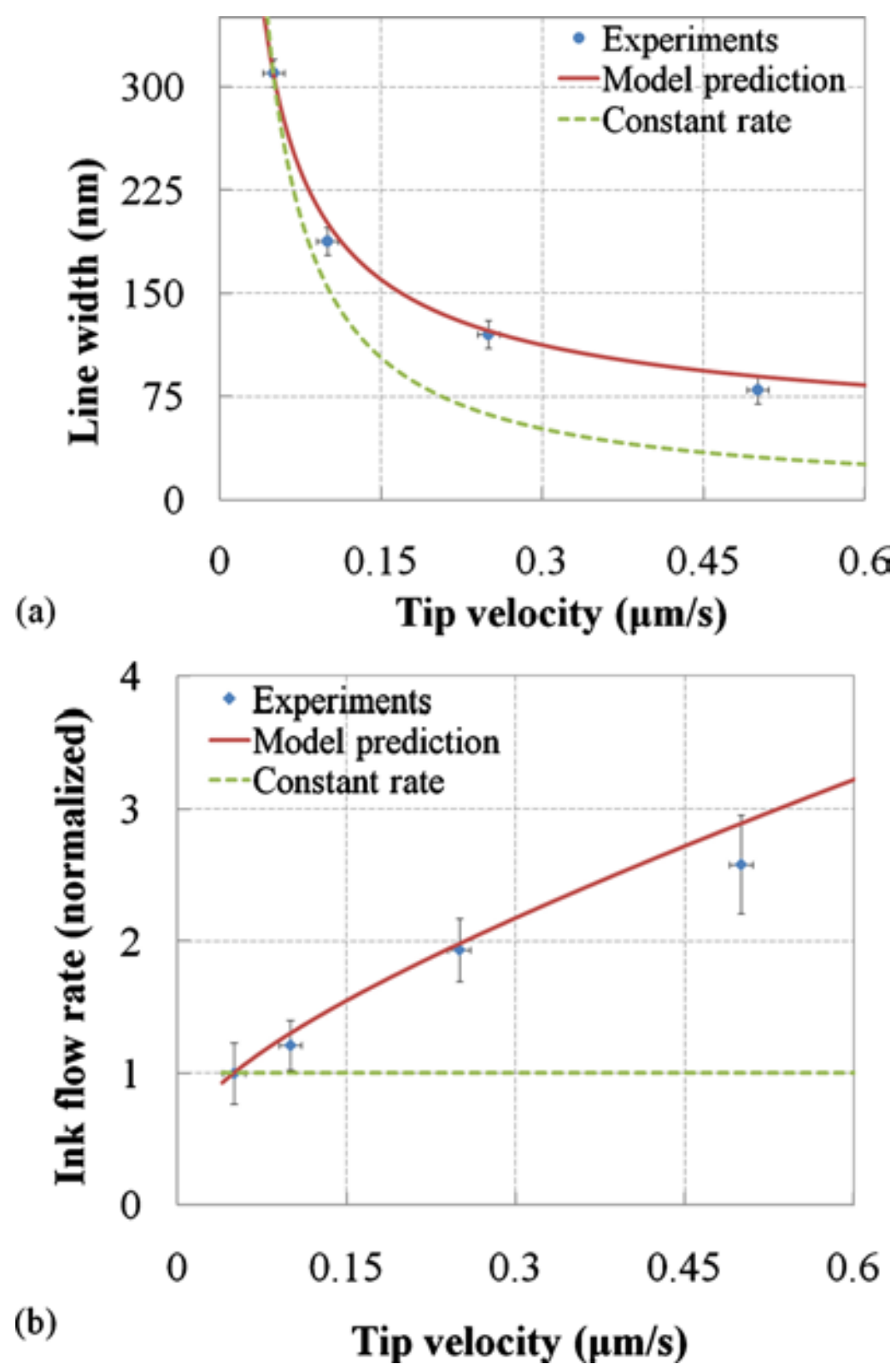

*Figure 2: Plot for moving tip (a) line width vs tip velocity (b) Ink flow rate vs tip velocity*

Another model mentioned in “Self-assembly of ink molecules in dip-pen nanolithography: A diffusion model” considers a continuum theory for the growth of circles.

$$D\frac{\partial P}{\partial r} = D'\frac{\partial P'}{\partial r}$$

Where P(r,t) is the density inside the circle and P’(r,t) is the density outside the circle. With some manipulations considering ‘n’ ink molecules per unit time are supplied, the final relation between the increasing radius and contact time is obtained as follows,

$$R(t)^2 = 4Dt\,\ln[(n\,/\,\pi\rho)\,/\,4D]$$

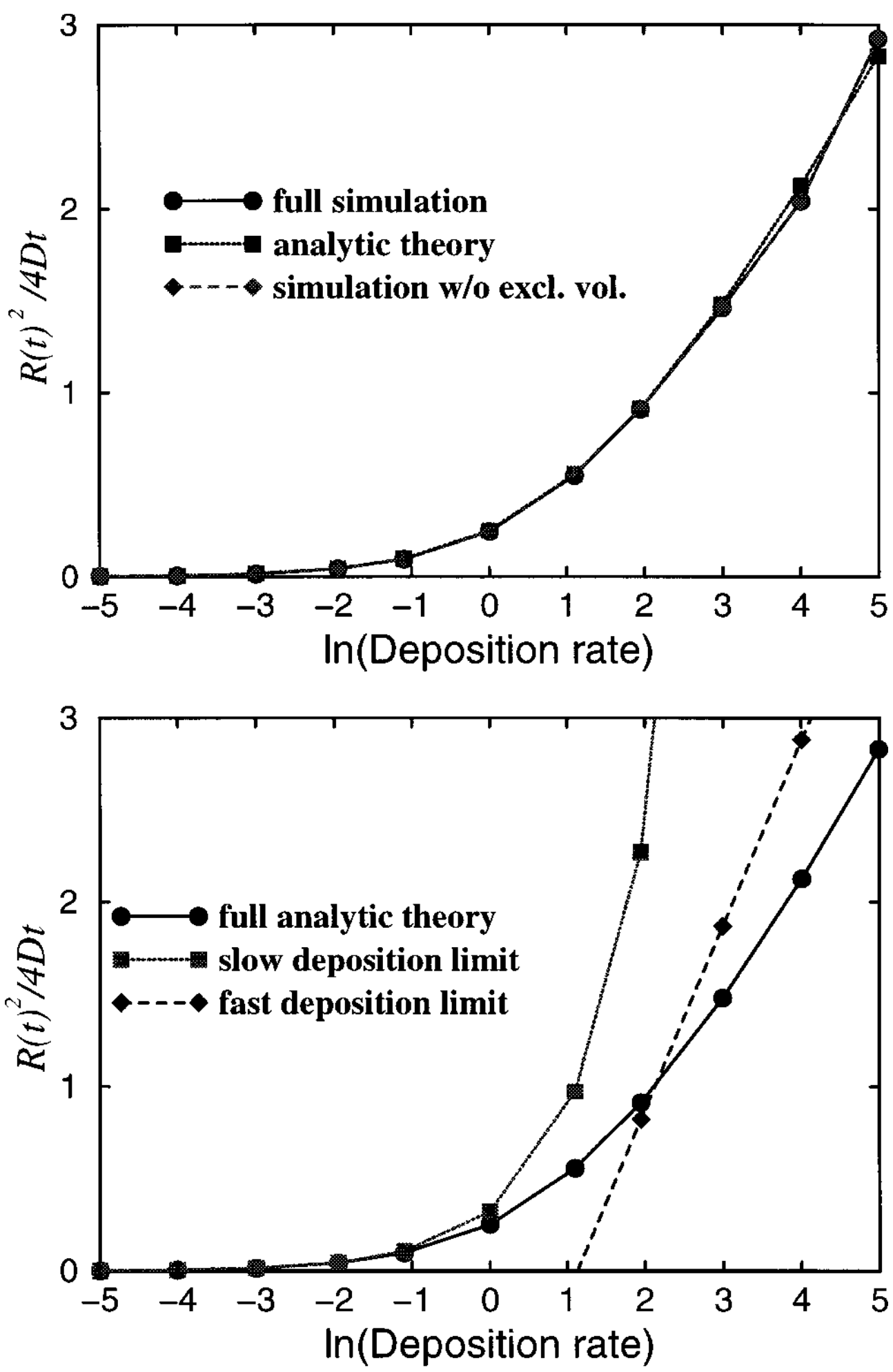

*Figure 3: Radius growth rate in time vs Deposition rate [2].*

The plots for the equation derived are shown in Figure 2. The Trend shows a negative curvature when radius of the ink drop is plot against time.

Another model which considers the molecular transport from the AFM tip to the substrate mentioned in "Molecular Transport from and Atomic Force Microscope Tip: A comparative study of Dip-Pen Nanolithography" shows a model,

$$F = \rho \Delta A \,/\, \Delta t = \rho w v$$

Where F is the flux of the molecules to the surface, ρ is monolayer density of the molecules and w is line width and v is the tip velocity.

Flux is also calculated for the circular drop as,

$$F = 2\pi r \frac{dC}{dr}$$

Combining the equations gives a relation between the constant terms and the variable terms as follows,

$$G = \left[{}^{4\pi rDC}/_{\rho}\right]$$

Where G is the product of all the constant terms [4].

A model mentioned in “Thiol Diffusion and the role of Humidity in Dip-Pen Nanolithography” [5] considers an equation for concentration with respect to time, and radius of the drop as follows,

$$C(r,t) = C_0 \frac{E_1(r^2 \,/\, 4D^t)}{\ln(4Dt \,/\, a^2 e^2 \gamma)}$$

The outcome plot for radius vs time is mentioned in Figure 4

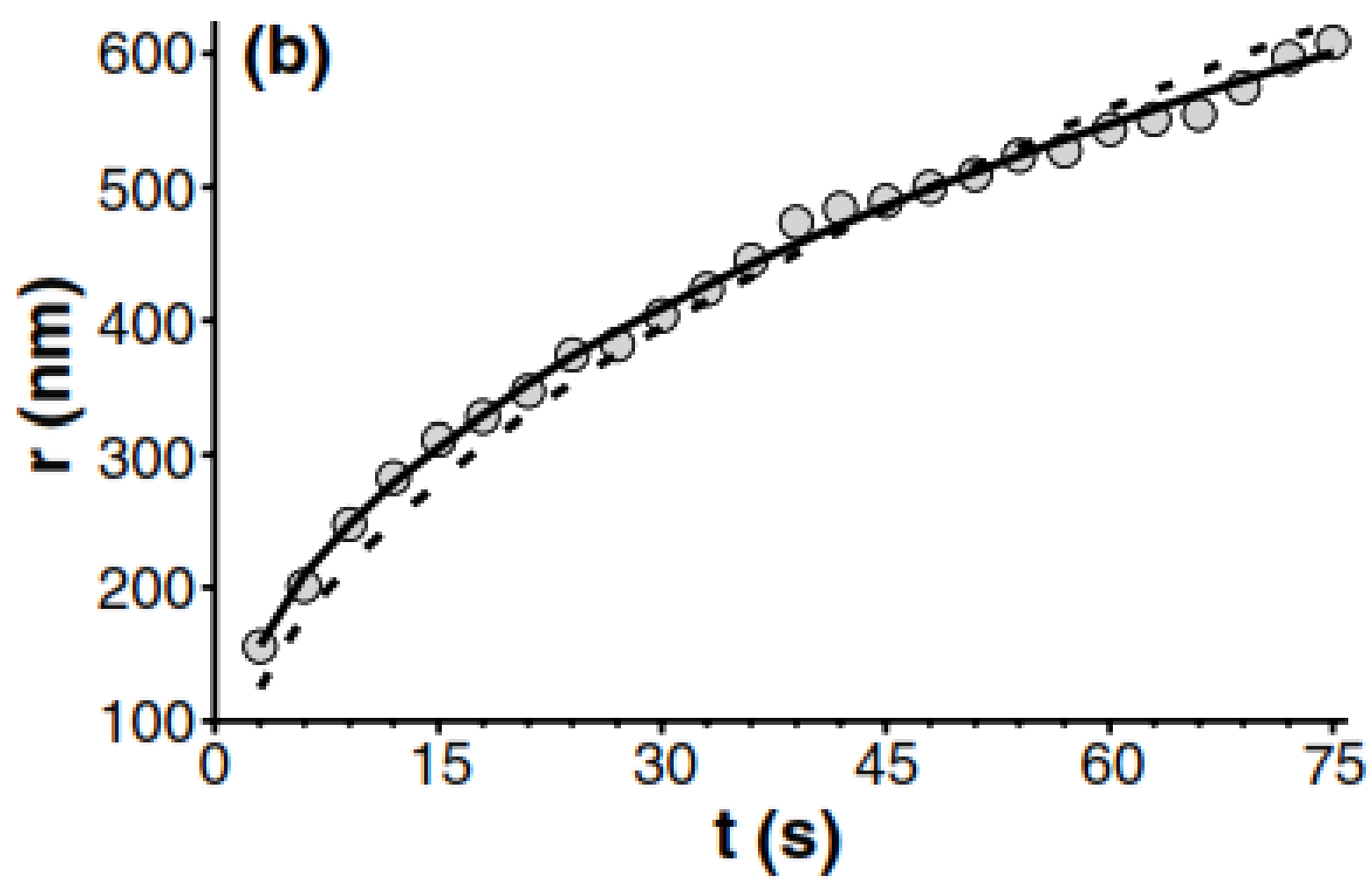

*Figure 4: Radius vs time plot for thiol diffusion on a gold substrate.*

A more complicated model for the ink transport is described in the research “A diffusive ink transport model for lipid dip pen nanolithography”. This model considers a two-step diffusion. The first step is the diffusion of the ink molecules from the tip of the AFM to the meniscus that forms around the tip. The second step is the diffusion from meniscus formed of the lipid ink to the substrate. To model this kind of diffusion process, many factors have to be considered such as, solubility of the ink, detachment rate of the ink molecules from the tip. The final equation of this model for the relation between all the factors is as follows,

$$\frac{8}{\pi}\frac{t}{\rho h} D_0 C_0 = \left[\frac{-B}{2\pi^z + D_s}\right] \omega^2 \ln\left(\frac{w}{2R}\right) + \left(1 + \left[\frac{\beta}{D_1 \pi R r_0 \,/\, L}\right]\right) w^2$$

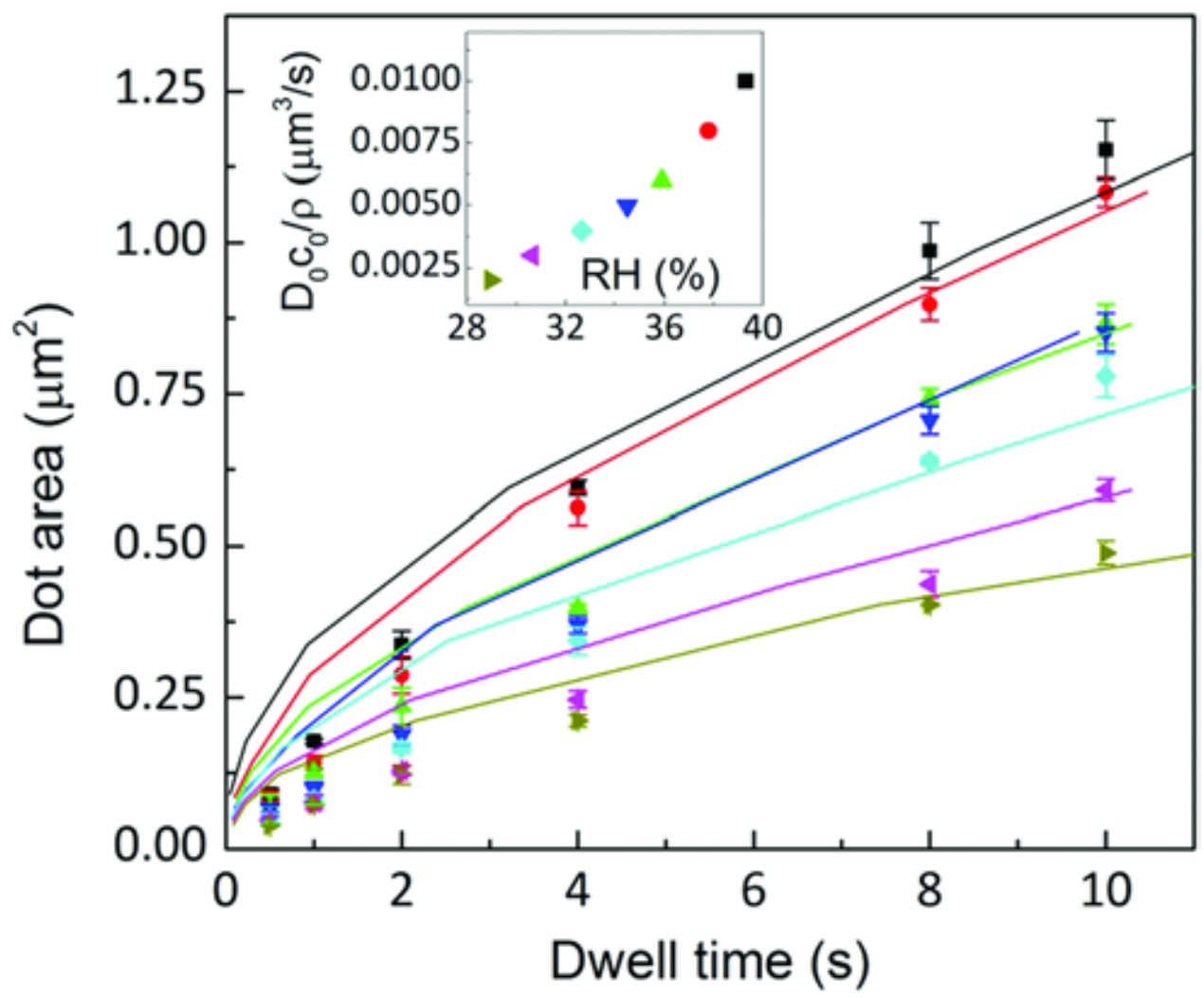

*Figure 5: Area of the ink dot on substrate vs Dwell time of the AFM tip [3]*

A comparatively simple model is developed in a research "Dip-pen nanolithography: A simple diffusion model" [6]. The model considers a steady state approach to simplify the diffusion model and a mass balance of ink molecules spreading with the radius and spreading with time gives an equation

$$t = \frac{C_m}{2DC_0}\left\{R^2\left[\ln\left(R/r_0\right) - \frac{1}{2}\right] + \frac{r_0^2}{2}\right\}$$

Where R is the advancing radius of the ink drop, $r_0$ is the initial radius of the ink drop and $C_0$ is the initial concentration of the ink drop. The relative plot of radius of the ink drop vs the time is as follows,

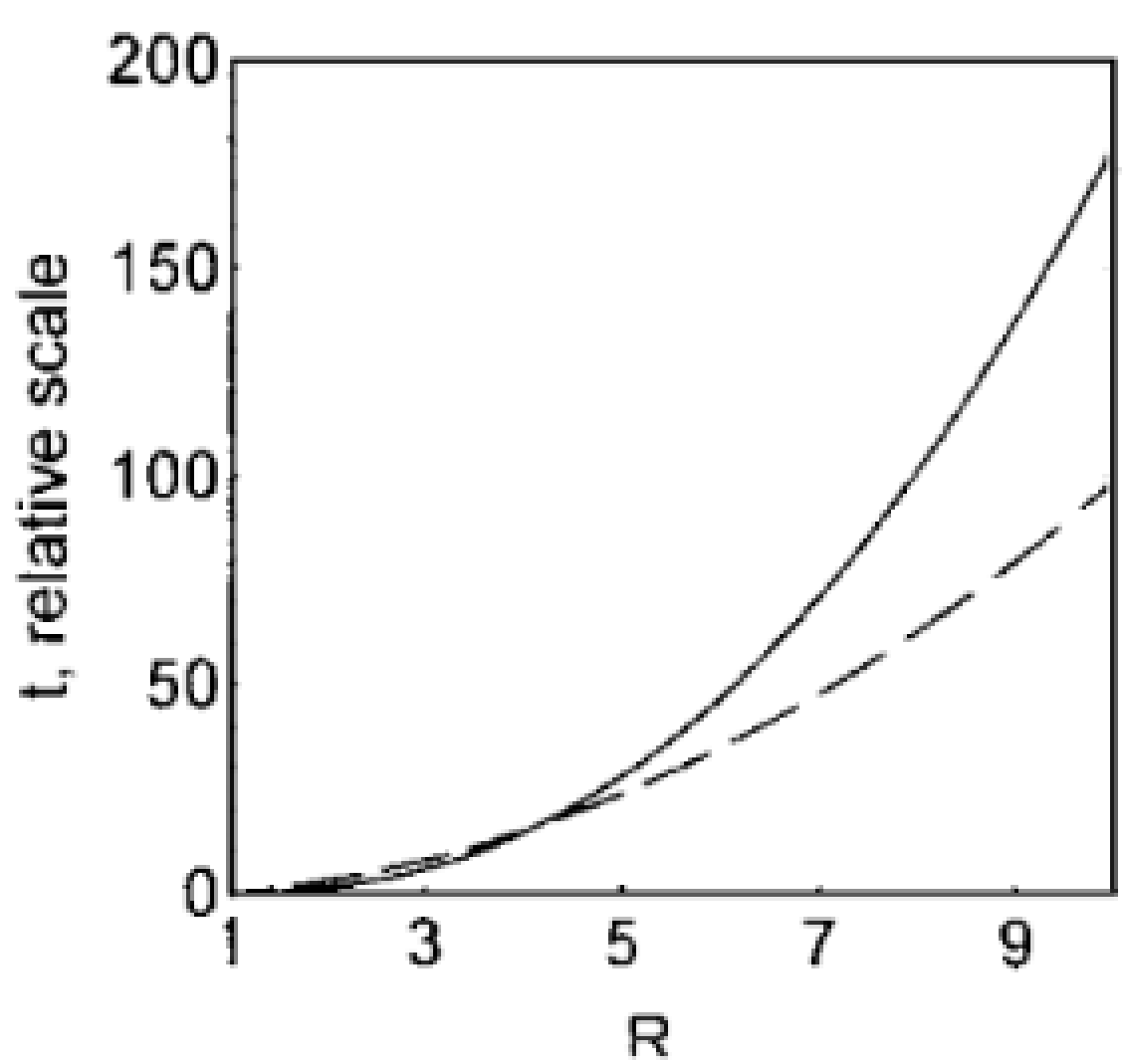

*Figure 6: Relative plot of radius of the drop to the time [6]*

## The new developed model

The aim of this paper is to develop is more simple diffusion model for the ink drop which will work for any system similar to dip-pen nanolithography which consists of an AFM ink tip, a molecular ink and considering a monolayer deposition of the ink drop on the substrate.

The model starts with a simple equation for the Fick's law for spherical co-ordinate system

$$\frac{\partial c}{\partial t} = \frac{1}{r}\frac{\partial}{\partial \sigma}\left[rD\frac{\partial c}{\partial r}\right]$$

Now the boundary conditions for the problem are specified,

1. The problem is considered steady state. Therefore, $\frac{\partial c}{\partial t} = 0$
2. The second boundary condition that is to be specified is the initial boundary condition. At the initial radius r = $r_{0,}$ the concentration has some value, i.e. C = $C_0$.
3. The next boundary condition is that the drop is considered monolayer once it transports to the substrate.
4. The last boundary condition is that to stabilize the drop the concentration at the interface at the end of spread is zero. Therefore, C(r(t)) = 0

The solution for the equation is given as follows,

$$0 = \frac{\partial}{\partial r}\left[r\frac{\partial c}{\partial v}\right]$$

Integrating with respect to r, we get;

$$C_1 = r\frac{\partial C}{\partial r}$$

$$\frac{c_1}{r} = \frac{\partial c}{\partial r}$$

Integrating again with respect to r, we get;

$$c_1 \ln r + c_2 = C(r)$$

Applying the initial boundary condition,

$$c_1 \ln r_0 + c_2 = C_0$$

Applying the final boundary condition,

$$c_1 \ln R + c_2 = 0$$

Therefore,

$$c_2 = -c_1 \ln R$$

Adding both the boundary conditions, we get;

$$c_1[\ln r_0 - \ln R] = C_0$$

Now, we calculate the values of the constants of integration $c_1$ and $c_2$,

$$c_1 \ln\left({}^{r_0}/_{R}\right) = C_0$$

$$c_1 = \frac{C_0}{\ln\left({}^{r_0}/_{R}\right)}$$

$$c_2 = \left[-\frac{C_0}{\ln\left(r_{0/R}\right)}\right] \ln R$$

Combining the equations, we get the solution for concentration with respect to the radius of the drop.

$$C(r) = \frac{C_0}{\ln\left(r_{0/R}\right)} \ln\left(\frac{r}{R}\right)$$

From the concentration of equation, we can derive the equation of the flux of the ink molecules for the circular spread,

$$J = -2\pi r D \frac{\partial C}{\partial r}$$

$$J = \frac{-2\pi r D C_0}{\ln\left(r_{0/R}\right)}$$

$$J = \frac{2\pi r D C_0}{\ln\left({}^{R}/_{r_0}\right)}$$

Now that we have got the equation of flux, there are two strategies we can use to relate the radius of the drop to the time.

1. Use the mass balance from Stefan Condition to relate the change in radius to the change in time. Therefore,

$$\text{Flux} = \text{Velocity} \times \Delta\text{C}$$

$$\frac{2\pi D C_0}{\ln\left({}^{R}/_{r_0}\right)} = r\frac{\partial r}{\partial t} \times \Delta C$$

Where,

$r\frac{\partial r}{\partial t}$ is the flux for a cylindrical system.

$\Delta C$ is the difference between the concentrations inside and outside of the interface, i.e. the edge of the drop.

Now, here the concentration inside the drop considering the initial condition is $C_0$ and the concentration outside the ink drop is zero as there is no ink. The equation becomes,

$$\partial t = \frac{\ln(R / r_0)}{2\pi D}$$

Now, Integrating the from r = $r_0$ to r =R, we can solve for R vs t. For the simplicity sake we will consider $r_0$ =1 and diffusion coefficient D = 1, we get;

$$t = \frac{(-1 + R)\ln(R)}{2\pi D}$$

We plot time vs radius with a relative scale, we get the plot as follows

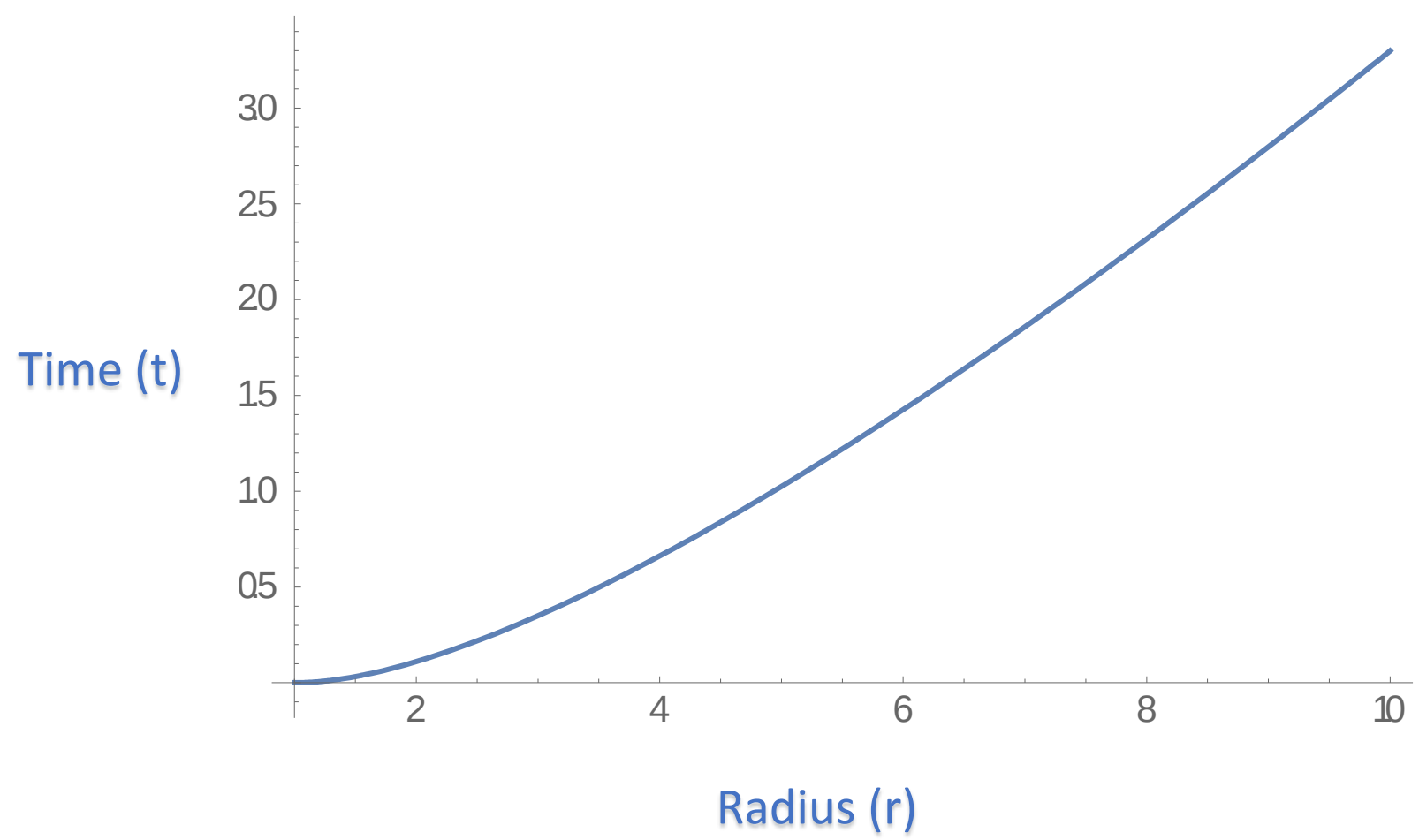

*Figure 7: Plot of Time of spreading vs Radius of the drop on a relative scale using mass balance theory*

Here, the Radius is plot on the X-axis and time is plot on the Y-axis. The plot is similar to plots plot of radius vs time shown in the previous researches.

2. The second strategy is to use the current equation to get the relation between the radius and time.

$$\text{Current} = \text{Flux} \times \text{Area}$$

Where,

Current can be defined in two ways

i. Number of atoms/molecules per unit time flowing/ arriving.
ii. Change in volume per unit time.

Now consider $n_a$ as number of atoms in transferred from the tip. And the as the drop extends outwards the area in this case will be the perimeter of the extending circle i.e. $2\pi r$. The equation becomes,

$$\frac{n_a}{\partial t} = \frac{2\pi D C_0}{\ln(R / r_0)} \times 2\pi r$$

$$\partial t = \frac{n_a \ln\left({}^{R}\!/_{r_0}\right)}{(2\pi r)^2 D C_0} \partial r$$

Integrating, we get;

$$t = n_a \times (-1 + R)\ln(R)$$

If we plot time vs radius again with a relative scale, we get the plot as follows,

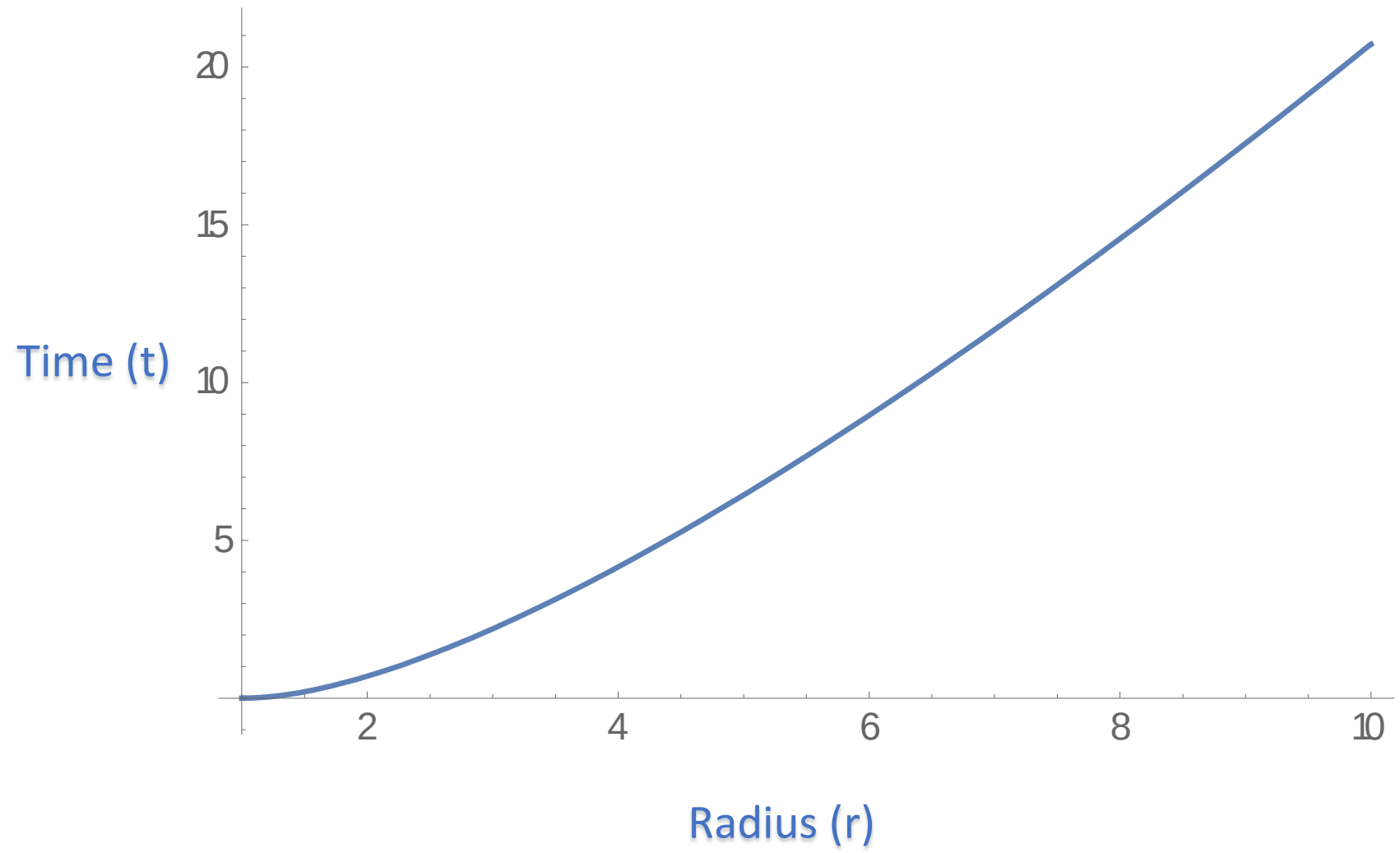

*Figure 8: Plot of Time of spreading vs Radius of the drop on a relative scale using current theory*

Here, the Radius is plot on the X-axis and time is plot on the Y-axis. This plot is also similar to the plots for radius vs time for the ink drop spread on the substrate in the mentioned researches.

The definition for current which says change in volume with respect to time can be also used to solve this equation. The boundary condition which states that the drop is monolayer conflicts with the calculation of the volume of the ink drop. If the volume is considered cylindrical, the integration has conditional solutions showing a slightly different profile.

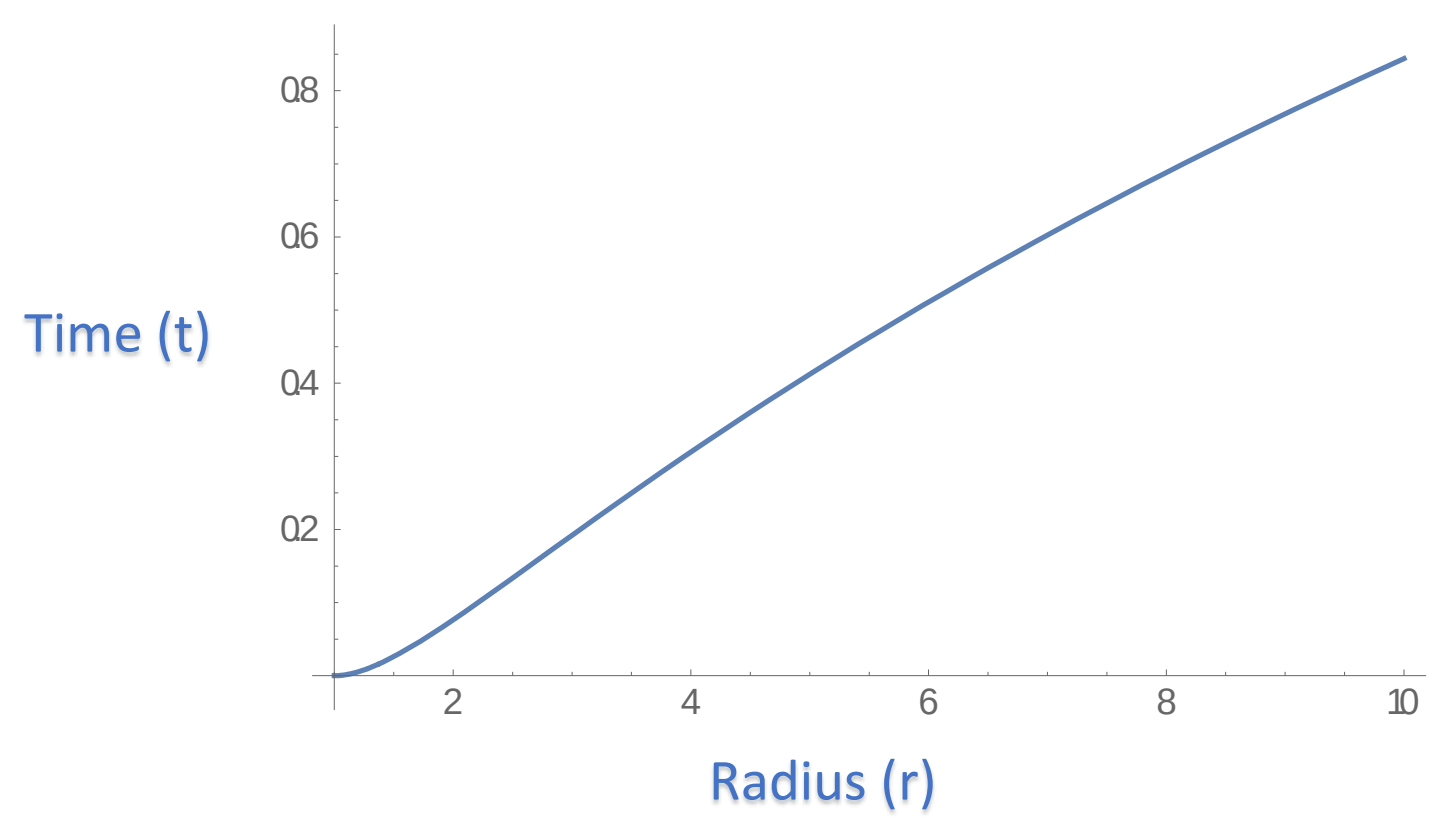

*Figure 9: Plot of Time of spreading vs Radius of the drop on a relative scale using current theory*

More work is needed to solve the diffusion problem with the mentioned strategy.

Since the these are relative plots for the equations, they are dimensionless, but the plots follow the same negative curvature profile that is expected for the growth of the ink drop on to the substrate.

## Conclusion

When we see the growth of the ink drop on the substrate, the growth of the ink drop, i.e. the radius increases linearly or with a slight negative curvature with respect to time. The paper tries to discuss a simple solution for the diffusion problem as how the diffusion problem can be modeled considering the most basic information needed without involving complicated factors. The solution might not be exact are very few factors are considered. But it might be considered as a preliminary guess or start of the diffusion problem with this kind of system. A detailed model can be developed when tackling any specific system such as when the type of ink, type of substrate is mentioned, the diffusion coefficient of the ink is known etc. The solution is supposed to work with all the system consisting molecular inks.